\documentclass{elsart}
\usepackage{graphicx,natbib,amssymb,lineno}

\journal{New Astronomy}

\begin{document}
\begin{frontmatter}
\title{Superdisks and the structural asymmetry of radio galaxies}

\author[NCRA]{Gopal-Krishna}
\ead{krishna@ncra.tifr.res.in}
\author[GSU]{Paul J.\ Wiita\corauthref{cor}\thanksref{nsf}}
\corauth[cor]{Corresponding author.}
\thanks[nsf]{PJW acknowledges support from a sub-contract to GSU 
from NSF grant AST-0507529 to the University of Washington.}
\ead{wiita@chara.gsu.edu}

\address[NCRA]{National Centre for Radio Astrophysics, TIFR, Post Bag 3, Pune University Campus, 411007, India}
\address[GSU]{Department of Physics and Astronomy, Georgia State
University, Atlanta, GA, 30302-4106, USA}

\begin{abstract}
We present a sample of 16 radio galaxies, each of which is characterized 
by a wide,   elongated emission gap with   fairly sharp and straight edges 
between the two radio lobes.    This particular subset of the ``superdisk'' 
radio galaxies is chosen because of a highly asymmetric location of the 
host elliptical galaxy relative to the gap's central axis. In addition to
posing a considerable challenge to the existing models, such a 
morphology also means that the two jets traverse highly unequal distances 
through the superdisk material. One thus has a possibility to directly 
investigate if the marked asymmetry between the two jets' interaction with the 
(much denser) ambient medium, during their propagation, has a significant 
import for the brightness of the hotspot forming near each jet's extremity. 
We also propose a new explanation for the formation of superdisks through 
the merger of a smaller elliptical galaxy with the massive host, in which 
the gas attached to the infalling galaxy deposits its angular momentum 
into the host's circumgalactic gas, thereby causing it to flatten into a 
fat pancake, or superdisk.   The asymmetric location of the host galaxy can 
be assisted by the kick imparted to it during the merger. We also suggest 
a physical link between  these radio galaxies and those with {\bf X}-shaped and 
{\bf Z}-symmetric radio lobes, commonly believed to arise from mergers of two 
galactic nuclei, each harboring a supermassive black hole.
\end{abstract}

\begin{keyword}
galaxies: active \sep galaxies: ISM \sep galaxies: jets \sep  radio: continuum
\PACS 98.54Ep, 98.54-w, 98.58.Fd, 98.65.Fz, 98.40.Dk
\end{keyword}
\end{frontmatter}

\section{Introduction}

A few years ago we noted  that a fraction of extragalactic double radio 
sources exhibit radio morphologies and other properties that led us to infer 
that they required gaseous ``superdisks" extending tens of kiloparsecs in 
both diameter and thickness \citep[hereafter GKW00]{gkw00}.
The key morphological property shared by these radio galaxies is the sharp, 
quasi-linear edges of the radio lobes on the sides facing the central
elliptical host galaxy \citep{gkw96,gkna},
as seen in at least a dozen radio galaxies at low to moderate 
redshifts.  We showed that such superdisks (SDs) can provide consistent,
alternative explanations for some key correlations found among the parameters 
of extragalactic radio sources, such as the Laing-Garrington 
effect \citep[e.g.,][]{garr88,lain}
and the correlated radio-optical 
asymmetries  \citep[e.g.,][]{mcca}.

  Furthermore, in high-$z$ RGs the strong tendency for redshifted component 
of the diffuse, quiescent Ly$\alpha$ emission to be on the side of the
brighter hot spot has been explained by postulating that the redshifted
Ly$\alpha$ emission originates from the region of the radio lobe on the 
near side of the nucleus, and is thus subject to much less dilution due
to the intervening dust 
\citep[e.g.,][]{hump,gkw05}. 
As discussed in GKW00, these Ly$\alpha$ RGs
provide optical evidence for high-$z$ SDs.  
Because the sharp-edged morphology would only be noticed 
when the radio jets are oriented close to the plane of the sky, SDs are 
subject to a strong negative selection effect, and GKW00 argued that even 
though the observed cases are few, the phenomenon may not be so uncommon. 
In this paper we present an enlarged sample of radio galaxies (RGs) exhibiting
SDs and discuss, in particular, their properties related to structural
asymmetry.

The sharp emission gaps seen in double radio sources are most commonly 
attributed to a blocking of the backflowing radio plasma in the lobes
by a denser thermal plasma  associated with the parent elliptical galaxy
\citep[e.g.,][]{leahw,wiitg,blac}.
Such backflows are to be expected and simulations indicate that they 
will indeed be diverted by galactic ISM \citep[e.g.,][]{wiitn}.
This approach was an alternative to the original proposal by \citet{sche}
according to which the central emission gap arose from the 
pinching of the inner parts of the lobes by the higher gas pressure of the 
host galaxy.  Both of these mechanisms, however, seem incompatible with 
cases where the edges of the gaps are long and straight; they seem particularly 
incapable of explaining the highly asymmetric SDs where the host elliptical 
is found almost at one edge of the radio emission gap.  A third possible 
explanation for large gaps has recently been put forward by \citet{gerg};
they suggest that SDs could be carved out as two galaxy 
cores containing supermassive black holes (SMBHs) merge and the associated 
pair of jets undergoes a rapid precession during the later stages of the 
merger.  While this novel process might indeed create a gap, it too may have 
difficulty in explaining the   sharp inner edges of the lobes, particularly 
those of the 
highly asymmetric SDs we are exploring here.

Initially we proposed that the SD is primarily made of the interstellar 
medium bound to the RG itself, perhaps originating from the cool gas belonging 
to gas-rich disk galaxies previously captured by the giant elliptical host of 
the powerful RG \citep[e.g.,][]{stat}.
We argued that the tidal 
stretching and heating of that gas during the capture was sufficient to 
produce very large fat pancakes \citep{gkna,gkw00}.
Since in some extreme cases (e.g., 0114$-$476; Table 1) SDs were found to have 
widths running into several hundred kiloparsecs, an alternative possibility 
was also considered in GKW00, according to which at least some SDs trace the 
gaseous filaments of the ``cosmic web". 

In a recent paper, we have attempted to explain the occurrence of SDs 
in high-$z$ RGs which have yet to acquire a significant circumgalactic medium 
(CGM) or reside in an intracluster medium (ICM) \citep[ICM;][]{gkwj}.
We argued that the material forming the SD in such RGs likely 
arises from the nuclear wind which is believed to be associated with the 
AGN activity \citep[e.g.,][]{soke}.
We investigated the conditions 
under which this wind material would be squeezed between the two radio lobes 
into a pancake shaped SD. It was shown that for a wide range of reasonable wind
and jet parameters those jets launched within a few tens of Myr subsequent 
to the initiation of wind production could quickly catch up to the wind-blown
bubble and then the lobes growing outside the bubble could indeed squeeze the 
bubble in such a way as to produce SDs.  Such a squeezing, however, is 
unlikely in $z < 1$ RGs, which are usually surrounded by a significant amount
of X-ray emitting CGM with a pressure at least that of the radio lobes \citep[e.g.,][]{cros}.
Therefore, another mechanism seems to be 
required at least for the SDs in low $z$ RGs.  

In \S 2 we tabulate a significant number of additional RGs with highly 
asymmetric SDs.  We also briefly discuss some interesting morphological 
properties 
of these asymmetric SDs. We then propose in \S 3 a new mechanism based on 
a merger of an elliptical galaxy, also rich in hot gas, with the massive host 
elliptical. In this scenario the orbital angular momentum of the captured 
galaxy's gas will be gradually transferred to the CGM gas of the host galaxy,
thus transforming its quasi-spherical CGM halo into a fat pancake. This 
scenario ties in well with our explanation for the {\bf Z}-symmetries seen in 
some {\bf X}-shaped RGs (XRGs) \citep[hereafter GKBW03]{gkbw}.
Conclusions are summarized in \S 4.

\section{Radio Galaxies with Asymmetric Superdisks}

Previously we presented radio and optical evidences for SDs in a dozen 
powerful RGs (GKW00). In these objects the SDs are 
clearly traced by highly 
extended, sharp and quasi-linear inner edges of the radio lobes; these give 
rise to strip-like emission gaps between the lobe pair 
\citep[also see][]{gkw96,gkna}. 
The median width of those SDs 
was found to be $\sim 25$ kpc and the median length at least three times 
greater. 

One important RG correlation that is otherwise hard to explain is 
the lobe-depolarization asymmetry \citep{lain,garr88,garr91}.
The passage of radio photons through a 
very wide SD can naturally produce much greater depolarization for the more 
distant lobe, and this effect is  practically independent 
of the extent 
of the RG (Gopal-Krishna \& Nath 1997). 
The  previous explanation in terms of an essentially 
spherical magneto-ionic halo requires the unlikely condition that the size 
of that halo is always maintained to be within a factor of two of the total 
extent of the radio source (GKW00). The SD picture can also explain some 
meterwave flux variability via ``superluminal refractive scintillations" 
\citep{gkri,camp}; also see \citet{ferr}.

Another striking correlation exists between radio lobes and the extended  
optical emission line regions (EELRs), in the sense that the peak surface brightness 
of the EELR (on scales of tens of kpc) nearly always occurs on the side of 
the shorter of the two radio lobes \citep[e.g.,][]{mcca}.
Their original 
explanation of this strong effect invoked large scale density asymmetries
in the CGM about the RG; on the denser side, jet 
propagation would be slowed down and thermal emissivity of the EELR filaments 
enhanced through better confinement. Although conceivable, 
this mechanism cannot explain why the correlation should be nearly perfect 
(37 of 39 cases) even when the lobe size asymmetry is very minute. 
Nor can this 
\citet{mcca} 
interpretation explain why the
brighter EELR -- shorter lobe correlation is more pronounced 
for higher-$z$ RGs (where it is measured in 
the [O II] $\lambda$3727 line) than for lower-$z$ sources (measured in 
H$\alpha$). An alternative explanation of this correlation in terms of SDs is discussed
in 
\citet{gkna} and GKW00.  

Our first sample of 12 SDs indicated that the hot spots 
in the radio lobe pair are usually located more symmetrically about the SD's 
midplane than about the host galaxy (GKW00). This greater symmetry with respect 
to the SD than relative to the current location of the host galaxy was clearly 
the case for 9 of the 12 RGs, and there was only one clear counterexample.  
The most plausible explanation for this result is that the galaxy has moved 
considerably from the SD's midplane since the AGN phase began. Such a motion 
would induce an asymmetry that would also
work in the right direction to obscure the portion of the EELR which falls on 
the far side of the galaxy as marked by the current position of the radio core; 
furthermore, any dust 
in the SD would enhance the correlation in the blue (GKW00).  Here we
find that this trend for radio lobes to be more symmetric about the SD 
than about the radio nucleus is further strengthened in our enlarged sample of 
SDs.

In this paper we present an enlarged sample of 16 SDs, of which 6 are 
common 
to the set presented in GKW00. The selection criterion for the current 
sample is a pronounced asymmetry of
the SD relative to the host elliptical galaxy, such that the galaxy is
offset from the midplane of the SD by at least a quarter of the SD's
average width; for this reason we do not include the other six
RGs discussed in GKW00. While we do not claim the sample to be complete in any
sense, we do expect it to be unbiased (and representative) from the 
viewpoint of the asymmetry, since it was assembled from a fairly extensive 
search of published radio maps such that no clear cases of asymmetric
SDs were left out of this sample.
Note that the source 3C 16 represents an extreme case of asymmetry, since
the host galaxy lies completely outside the emission gap, within a radio 
lobe.  The SD case is strong, and although the offset of the host from the
center of the SD is large ($\sim 80$ kpc), this distance could have been 
covered by the host galaxy moving at $\sim$ 350 km s$^{-1}$ over an active 
lifetime of $\sim$250 Myr 
\citep[e.g.,][]{bara}.
Another notable
aspect of this sample (Table 1) is the presence of 3 double-double RGs 
\citep{lara99,scho},
namely 3C 16, 0114$-$476 and 1155$+$266 (Table 1).

For each RG, identified in columns 1 and 2 of Table 1, the redshift is 
given in column 3, while columns 4 and 5 give 
the largest angular size (LAS) of the RG and the width of the SD, respectively,
in arcseconds. Column 6 converts that SD width to kpc, 
using a cosmology with $H_0
= 70$ km s$^{-1}$ Mpc$^{-1}$, $\Omega_\Lambda = 0.7$ and $\Omega_m = 0.3$. 
References to all of the radio maps are
given in the last column of Table 1.

Contour maps (or radio photographs) of three of the RGs in our sample,
namely, 3C 227, A1425 and 3C 381, are shown in Figs.\ 1--3, where the host 
galaxy is seen to be situated close to one edge of the SD. These maps are 
taken from \citet{blac}, \citet{owen} and \citet{leahp},
respectively.

\begin{figure}
\includegraphics[scale=0.14]{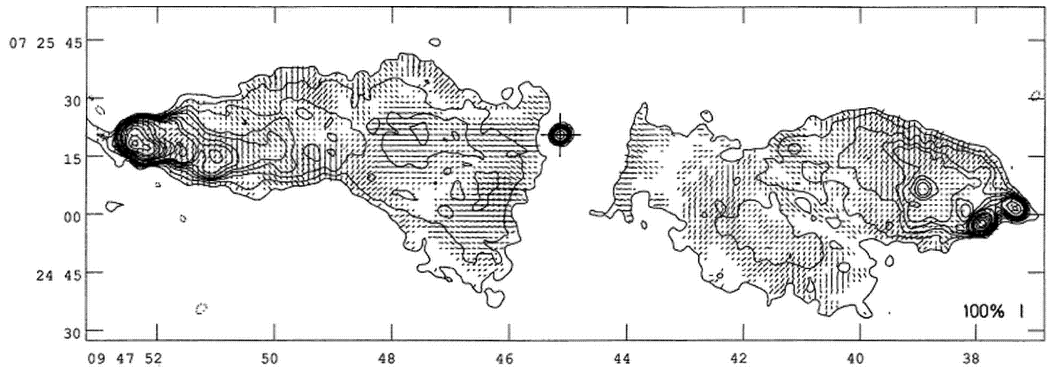}
\caption
{Radio map of 3C 227 (0945$+$076) from 
\citet{blac}. 
 Contours are shown at (-3, 3, 6, 9, 12, 15, 20, 25, 30, 50, 70, 100, 150, 200, 300,
400, 500)$\times$ 150 $\mu$Jy beam$^{-1}$.  Percentage polarization values (E vectors) are also shown.\copyright Royal Astronomical Society; reproduced by
permission of the RAS. }
\end{figure}

\begin{figure}
\includegraphics[scale=0.28]{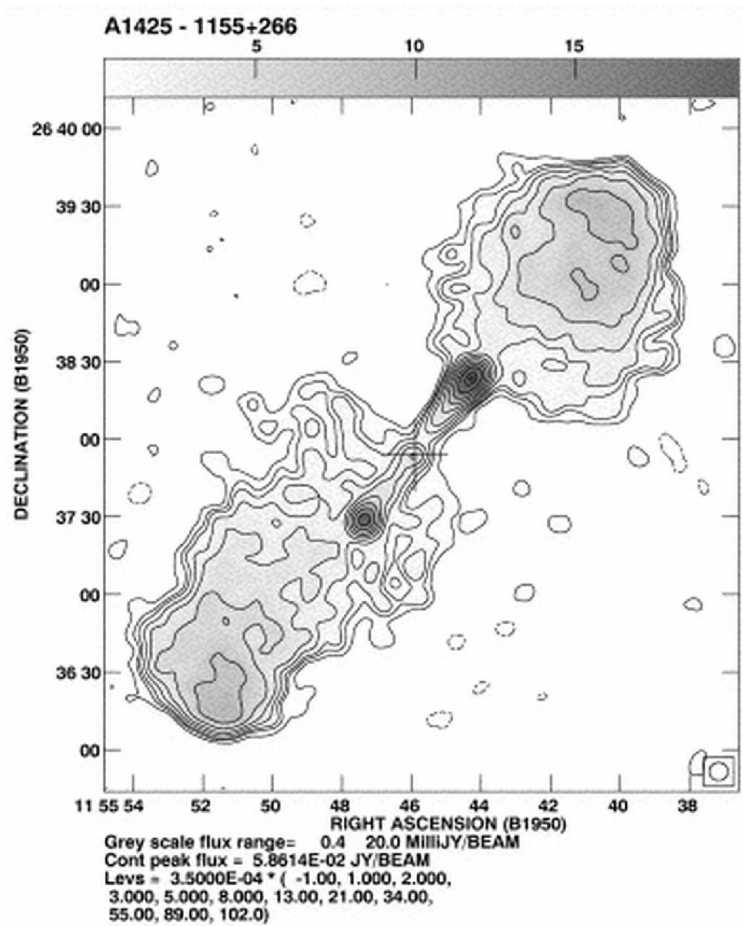} 
\caption
{Radio map of A1425 (1155$+$266) from 
\citet{owen};         
\copyright American Astronomical Society; reproduced by
permission of the AAS.
}
\end{figure}

\begin{figure}
\includegraphics[scale=0.28]{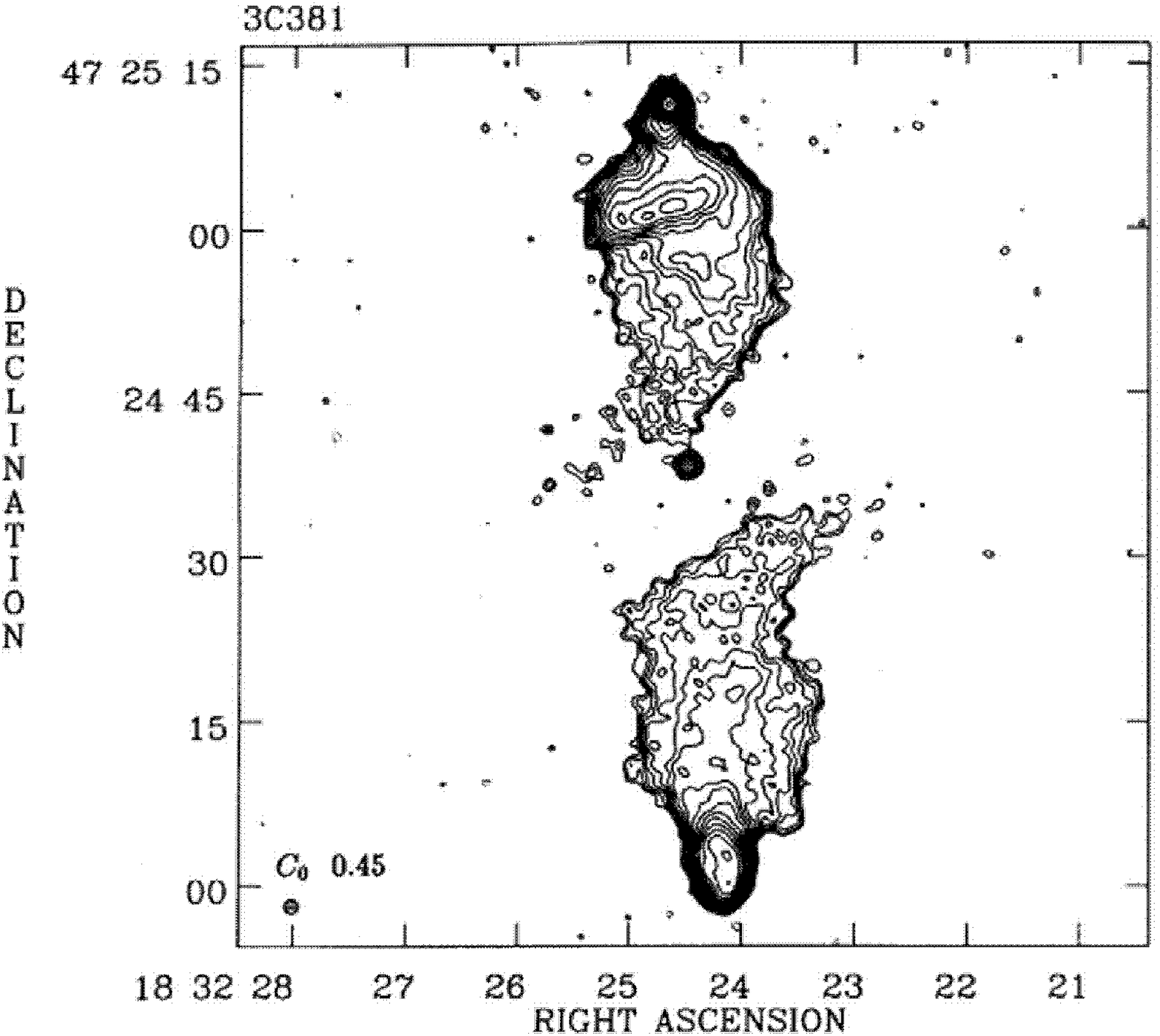} 
\caption
{Radio map of 3C 381 (1832$+$474) from 
\citet{leahp}.  
Contours are logarithmic with a ratio of $\sqrt 2$ between levels; i.e., levels are
at $C_0 \times 1, \sqrt2, 2, 2\sqrt2, 4, .\ .\ .\ )$.  The value of $C_0 = 0.45$mJy beam$^{-1}$ is given in the panel and
is about 3 times the rms noise level; the hatched circle indicates the FWHM restoring beam.   \copyright American Astronomical Society;  reproduced by
permission of the AAS. 
}
\end{figure}

In order to illustrate the way in which superdisks were identified 
and their properties measured, we show in Figs.\ 4--6 the outermost contour of 
the radio map  of six of our SDs (see Table 1 for map references). In one case, 
0114$-$476, a slightly higher contour is reproduced in order to illustrate 
the outline of the SD more clearly. The `+' sign marks the radio core 
at the nucleus of the host galaxy.  Also reproduced are one or two contours 
closely encircling the peak of the hot spot within each radio lobe. In 
the case of 0114$-$476, the inner double lobes are also detected and are shown.
The edges of each SD, as adopted here, are bracketed by 
a pair of straight lines and the central axis (i.e., mid-plane) of the SD 
is indicated with a broken line.
In all cases, the published radio contours allow us to delineate fairly well 
the two edges of the SDs, despite relatively narrow protrusions into the SD 
region which are seen in some of the RGs and are probably associated with 
the radio jets. 

\begin{figure}
\includegraphics[scale=0.52]{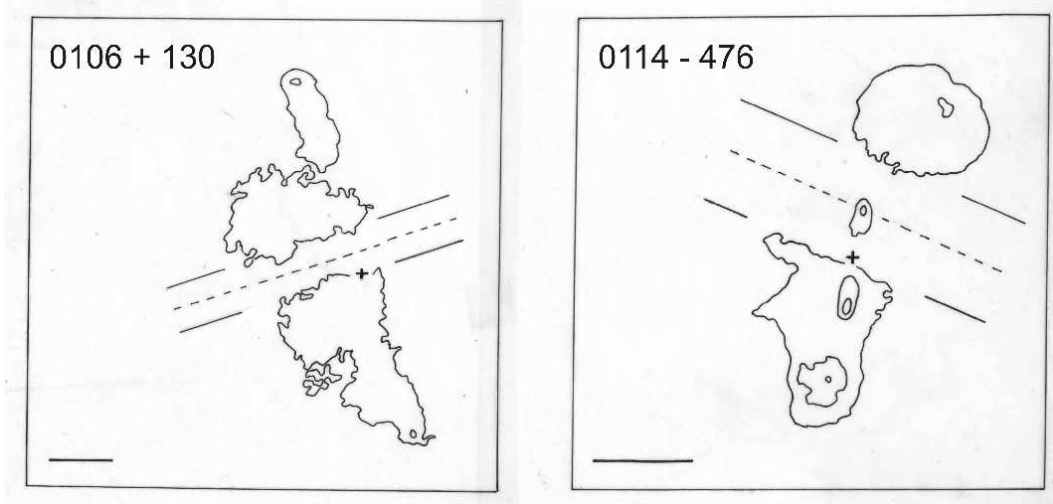} 
\caption
{Outlines of radio maps of 0106$+$130, 0114$-$476, 
with `+' signs giving the location of center of the host galaxies, 
 straight line segments illustrating the edges we adopt for the SD and dashed lines giving the
SD midplanes.  The hotspot location is noted for each lobe and 50 kpc scale bars are drawn in
each panel, except for the RG 0114$-$476, where it measures 500 kpc.
The maps from which these contours were traced are respectively from references 2 and 5 in Table 1.}
\end{figure}

\begin{figure}
\includegraphics[scale=0.52]{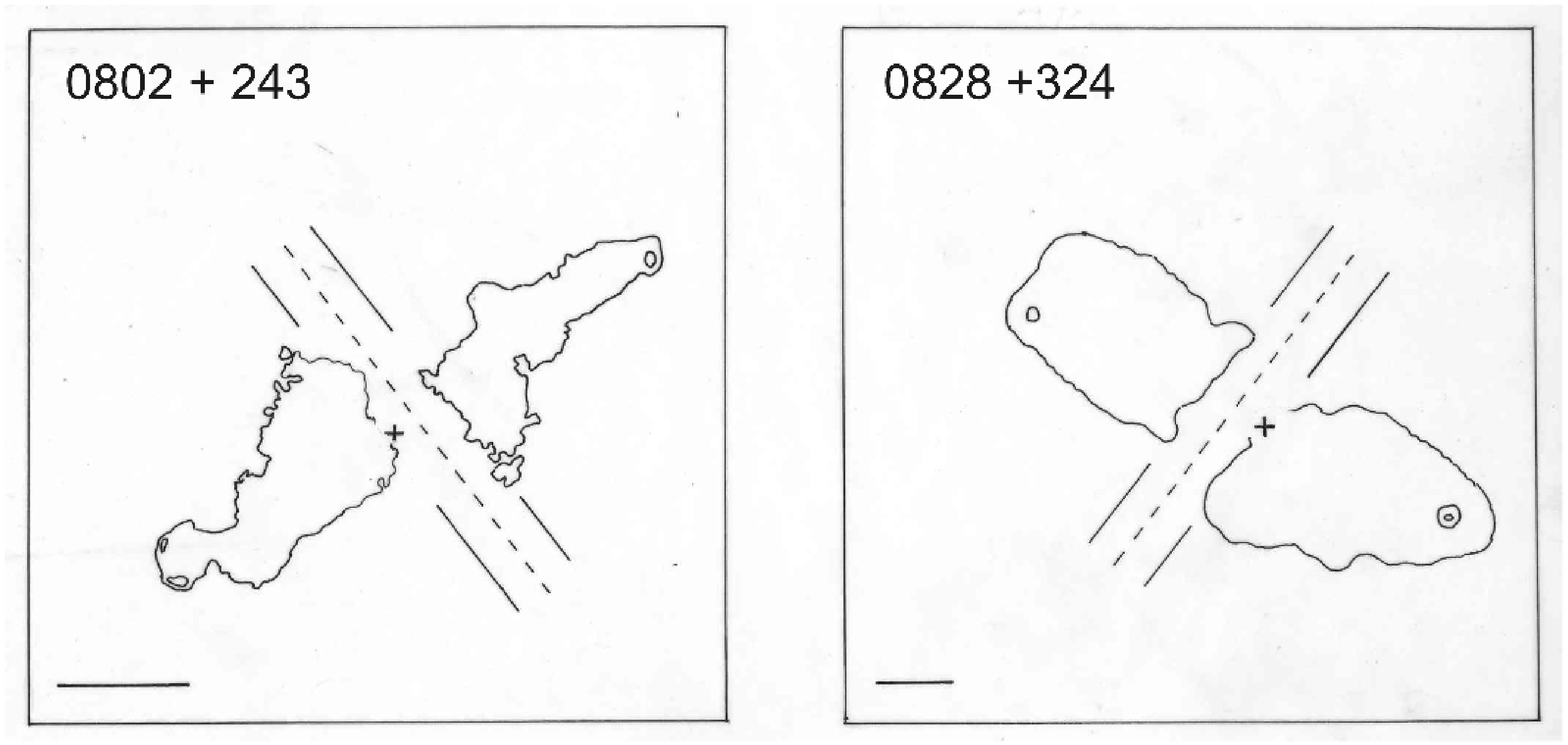} 
\caption
{As in Fig.\ 4 for the RGs 0802$+$243, 0828$+$324, from references
9 and 11.}
\end{figure}

\begin{figure}
\includegraphics[scale=0.52]{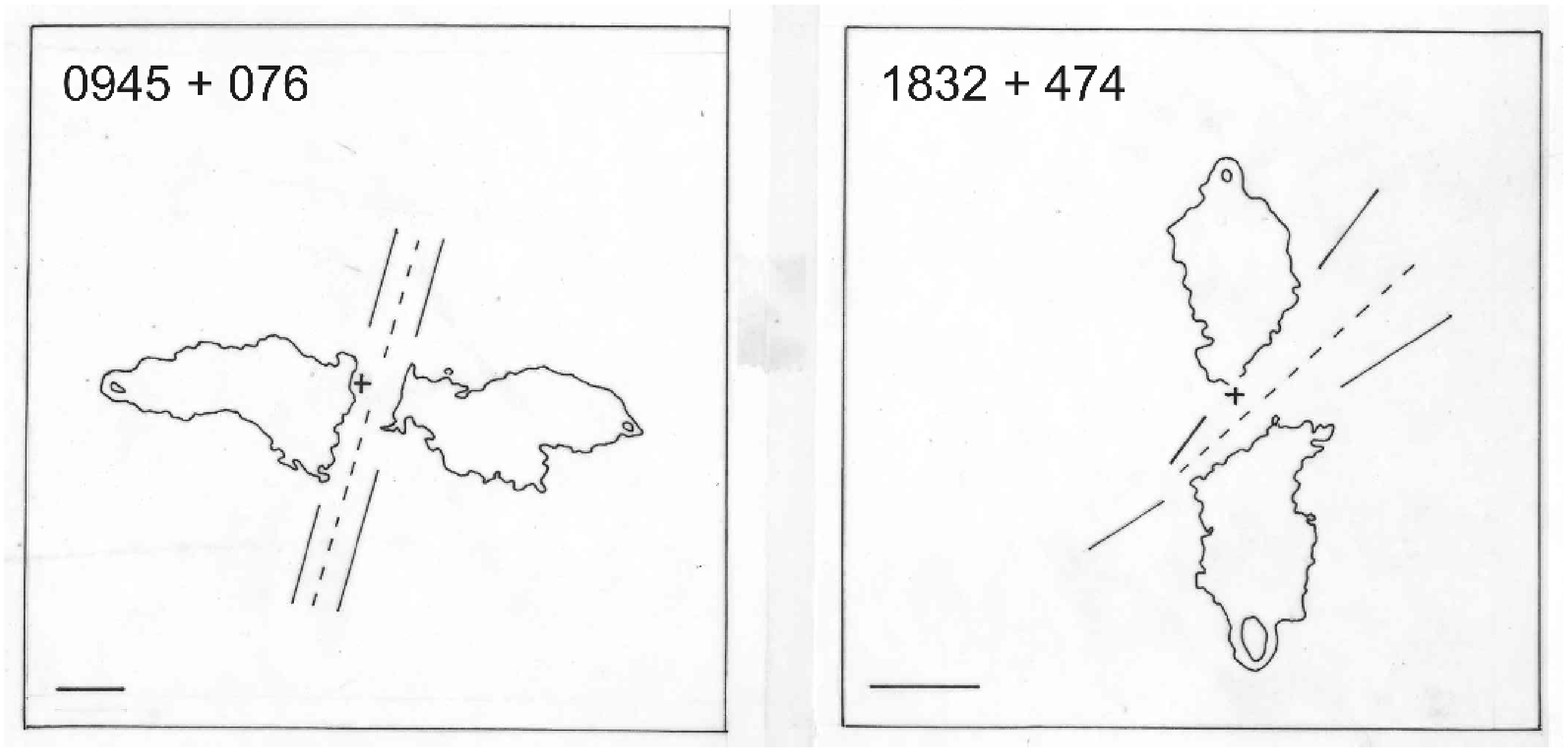} 
\caption
{As in Fig.\ 4 for the RGs 0945$+$076, and 1832$+$474, from references
12 and 2.}
\end{figure}

\begin{table}
\caption{Radio Galaxies with Asymmetric Superdisks}
\begin{tabular}{llccccccccccl} \hline
\noalign{\smallskip}
Source 	  &Other& z  &LAS  &~~SD &Width &R$^a$  & Q  &  Q$_{\rm sd}$ & SD Sym  &$\psi$&Fade & Refs \\
&  Name& &$^{\prime\prime}$&$^{\prime\prime}$&kpc&&&&&$^{\circ}$&\\
\noalign{\smallskip}
\hline
\noalign{\smallskip}

0035$+$130$^b$  &3C16 &    0.405  &73&  13 &  70 &S32 & 0.49  &1.82 &Y & 14 &Y &1,2 \\
0106$+$130$^b$  &3C33 &    0.060 & 70 &  8 & 9 & S5.7&0.82 & 1.08 & Y &2&Y &2 \\
0106$+$729  &3C33.1 &  0.181 & 55 & 19 & 58  & N6.4& 0.59 & 0.96 & Y& 0& Y &3,4\\
0114$-$476    &     &   0.146 & 635&  200& 510 & 1& 0.71 & 1.34 & Y & 20& ? &5\\
0117$-$156  &3C38&    0.565  & 13 &   3 & 19 & N2& 0.52 & 0.95 & Y& 11 &Y &6\\
0453$+$227  &3C132 &   0.214&   22&    4& 14 &W2 & 1.06 &1.23   &N & 2&Y & 7\\        
0528$+$064  &3C142.1& 0.406 &  49 &   6  &33 & W2 & 0.61  &0.52&N  &10 &N  &8 \\
0605$+$480  &3C153&   0.277&    7&    2 & 8 & E2 & 1.95  &2.52   &N & 1&N & 7 \\ 
0802$+$243$^b$  &3C192&   0.060 & 190 &  27& 31  &W4 &1.18   &1.02&Y& 5  &N &9,10  \\
0828$+$324$^b$  &4C32.25& 0.051&  307 &  52& 52 & W2.9&0.80   &1.09   &Y & 0&Y& 11  \\
0945$+$076$^b$  &3C227 &   0.086&  230&   23&37 & E2& 1.08  & 1.01  &Y & 10&Y &12   \\
1155$+$266  &A1425 &   0.140&  240&   50&123 &  1&0.92   &1.46   &N & 1&? &13   \\
1409$+$524  &3C295 &   0.464&  4.7&  1.7&10 & N1.8 &0.67   &1.12   &Y  & 0&Y  &14  \\
1832$+$474$^b$  &3C381 &   0.161&   71&   11&30 &N9  & 0.86  &1.05   &Y & 1&Y &2  \\
1939$+$605  &3C401 &   0.201&   22&    3&10  &1 & 0.70  &1.10   &Y & 15&? &7,15 \\
2104$-$256  &NGC7018& 0.039 & 112 &  35& 27& N1.4  & 0.67  &1.35   &Y &18 &Y &16 \\

\noalign{\smallskip}
\hline
\end{tabular}\\
\\
Notes: $^a$the brighter hotspot (E or W; N or S) and the ratio of its peak contour to the fainter's;
$^b$also in Table 2 of GKW00.
References: 1) \citet{harv}, 2) \citet{leahp}, 3) \citet{vanb}, 4) DRAGN atlas:
www.jb.man.ac.uk/atlas, 5) \citet{sari}, 6) \citet{morg99}, 7) \citet{hard},
8) \citet{boge}, 9) \citet{leah97}, 10) \citet{baum}, 11) \citet{cape},
12) \citet{blac}, 13) \citet{owen}, 14) \citet{perl}, 15) \citet{burn}, 16) \citet{morg93}
\end{table}

We first summarize some salient features of the present sample of 16 RGs.
They lie within a redshift range from $\sim$0.04 to $\sim$0.6 and are thus in 
the cosmic age 
regime where each should have acquired a substantial amount of hot circumgalactic gas. 
In all cases the optical counterpart is a galaxy; this
is expected since the sharp-edged gaps in the radio bridge are likely to be 
observed only when the jets are close to the sky plane 
\citep[a necessary 
condition for identification as a galaxy, and not a quasar, 
according to the unified scheme for radio powerful AGN, e.g.,][]{bart}. 
Another property consistent with the unified scheme is that kpc-scale jets 
are clearly visible in just a few of these RGs, since Doppler boosting will 
not be  significant for the jet emission. 

Projected linear sizes of these 16 RGs range between 28 and 1620 kpc 
while the widths of the central emission gaps are between $\sim$8 and 510 
kpc. Discarding these two extreme values, the mean of the other 14 projected
SD widths is 37 kpc and the median is 31 kpc. Nuclear radio cores are 
detected in 12 of the 16 RGs  
(see references in Table 1).

Column 7 of Table 1 identifies the brighter hot spot and also gives $R$, the ratio 
of the highest contour values for the two hotspots in the opposite lobes.
Column 8  gives the ratios, $Q$, of the distances of those two hot spots 
 from the radio core (or host galaxy).  The ratios, $Q_{\rm sd}$, 
of the distances of the hot spots from the mid-plane of the SD, as measured from 
the intersection of the radio axis (i.e., the line joining the two hot spots), 
are given in Column 9.  Here we define $Q$ and $Q_{\rm sd}$ so that they correspond to the 
distance to the brighter hotspot divided by the distance to the fainter hot 
spot. The selection of the brighter of the two hot spots  is done by comparing 
the highest contour values read from the 
radio maps cited in Table 1 and partly reproduced in the captions to Figs.\ 1--3.  
Defined in this manner, and based upon multiple measurements by
both authors, the standard errors on $Q$ are $\sim 0.01$; however, because of the
uncertainty in the exact location of the midplane of the SD, the errors
on $Q_{sd}$ are a bit larger, but still $\lesssim 0.03$.
Column 10 (``SD Sym") indicates 
whether the value of $Q_{\rm sd}$ (or its inverse) is closer to 1 than is 
$Q$, i.e., if the lobe   extents are more symmetrical with respect to the 
superdisk's midplane than relative 
to the nucleus.  Column 11 gives the misalignment angle, $\psi$, defined as 
the supplement to the angle between the lines connecting the two hotspots to the 
nucleus.  In column 12 (``Fade'') `Y' indicates when the jet traversing the 
greater distance through the SD produces the fainter of the two hotspots.

Fig.\ 7 gives a histogram of values of $Q_{sd}$ and $Q$ where we discard
the hot spot brightness ratio criterion and so define each value such that $Q \ge 1$; i.e., if 
the value in Table 1 is $< 1$ we replace it by its inverse.
As noted for a smaller sample in GKW00, the values of $Q_{\rm sd}$ are 
usually closer to 1 than are those of $Q$.  The present larger sample
strengthens this trend; the probability of having 12 (or greater) Y's in 
Column 10 by chance is only 0.038. This highly significant correlation is 
consistent 
with the scenario that at the time of jet launching the RGs were fairly 
centrally situated within their respective SDs and have gradually drifted 
outward since then (see GKW00).

\begin{figure}

\includegraphics[scale=0.60]{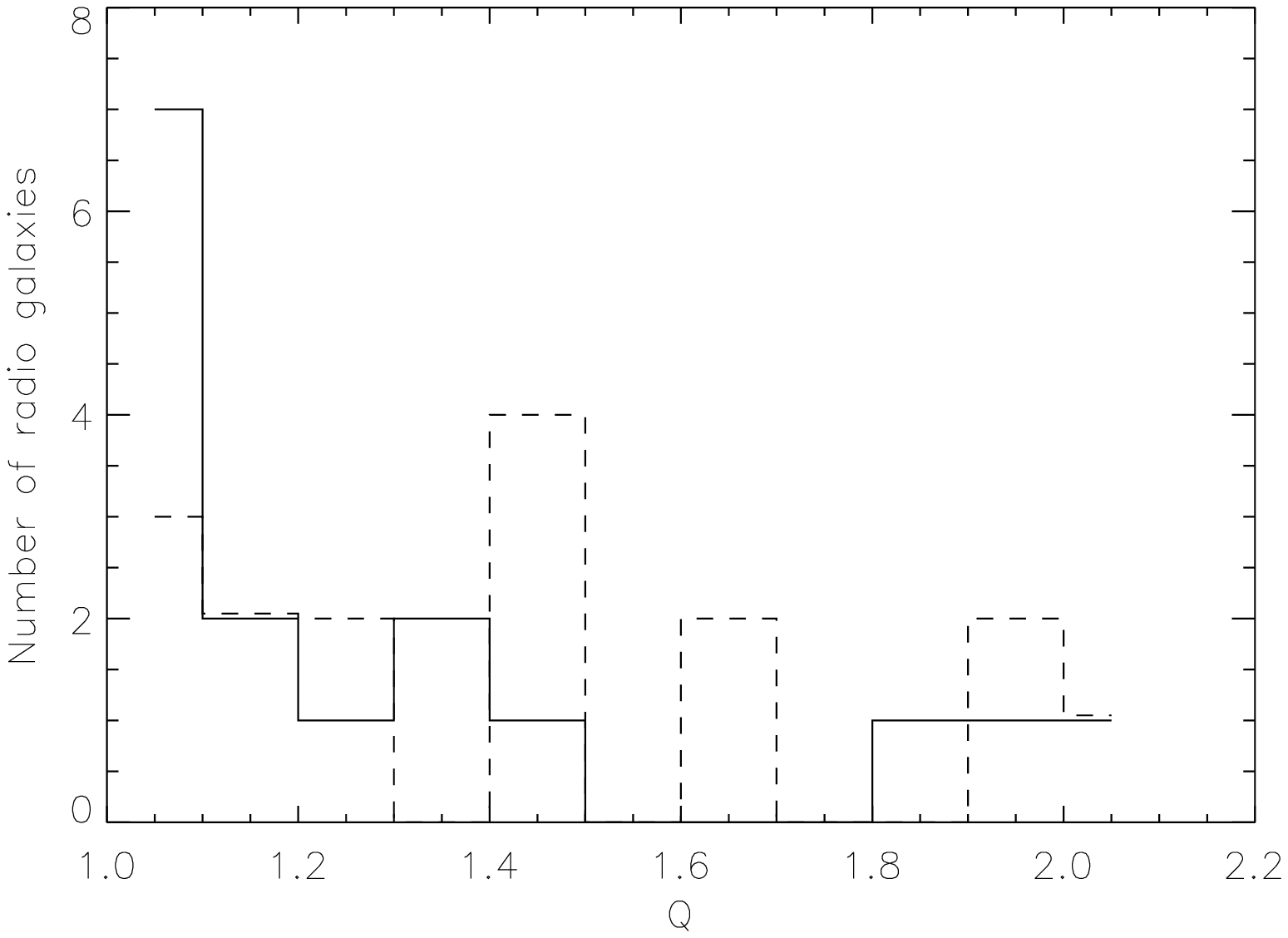} 
\caption
{Histograms of the distributions of $Q_{\rm sd}$ (solid lines) and
$Q$ (dashed lines); the last bin is comprised of $Q_{\rm sd} = 2.50$ along with
$Q = 2.04$.}
\end{figure}


Another suggestive correlation is noted for the cases where the sense of 
$Q$ and $Q_{\rm sd}$ are found to be opposite, with one ratio, nearly always 
$Q$, less than one, while $Q_{\rm sd} > 1$.  In every case where 
$\psi \ge 14^{\circ}$, i.e., 3C 16, 0114$-$476, 3C 401 and NGC 7018, this 
ratio is inverted, though such a trend is also found for a few cases with 
smaller $\psi$ (i.e., 3C 33, 4C32.25, and 3C 381, Table 1). Thus, while not
a strong correlation, 
the tendency of the double radio sources with highly misaligned  hotspots (i.e.,
larger $\psi$) to also have clearly  flipped values of $Q$ and $Q_{\rm sd}$ is consistent 
with the notion of the host galaxy moving relative to the SD, as inferred above.

A  stronger correlation found in the present sample
is for the fainter of the two hotspots to be found at the termination of the 
jet that
travels a greater distances through the SD. In 13 of the 16 RGs, the brighter of
the two hot spots can be unambiguously identified.  Two of the three RGs
where the fluxes are nearly equally bright, 0114$-$476 and A1425, 
are double-double sources, where the weakened outer hot spots are probably
not being currently fed 
\citep{scho}. 
Of these 13 sources with clearly unequal hot-spot fluxes,  10 show a fainter hotspot to be associated with jet traversing
a  longer column of the SD material (column 12 of Table 1). 
The probability of   10 or more 
out of 13 cases adhering to this trend is only 0.046.
Such a behavior can be explained by a stronger interaction of the jet traversing
the denser SD material,  conceivably causing a
larger entrainment of that material into the jet, as compared to the
jet propagating mostly through the much lighter lobe plasma.  

\section{Rotating the Circumgalactic Medium}

To produce a powerful RG a massive galaxy is required \citep[e.g.,][]{jahn}
and therefore at least one significant merger (defined as a
merger
one where the ratio of masses exceeds 0.1) is to be expected in the life of 
that galaxy  \citep[e.g.,][]{park}.
Such a merger, via spinning up the resultant 
black hole, may trigger the jet formation \citep[e.g.,][]{wils},
although it is now becoming clear that not all powerful RGs show clear signs of major mergers, i.e., those where the galaxy mass ratio exceeds 0.3 
\citep{tadh}, and so such mergers are  not required for incitement of RG activity.
The usual environment for powerful RGs is a group or sparsely populated
cluster of galaxies \citep[e.g.,][]{wold}.
To check the feasibility of the angular 
momentum transfer concept put forward in Sect.\ 1, 
we carry out a rough estimation using reasonable 
fiducial values for the various parameters. So we consider a massive 
elliptical host galaxy, of baryonic mass $M_h \simeq 5 \times 10^{11}M_{\odot}$, 
capturing a smaller, but still significant, elliptical group member, of mass 
$M_s \gtrsim 1 \times 10^{11}M_{\odot}$.  This galaxy is taken to initially have
a mass, $M_{g} = f M_s$, of hot gas, with $f \lesssim 0.1$. The larger galaxy
will possess a hot halo of gas, which we shall denote as its CGM, with typical core radius, $a \approx 1$ kpc, 
central density $n_0 \approx 10^{-1}$ cm$^{-3}$, and density distribution 
$n_{CGM}(r) \approx n_0/[1 + (r/a)^2]^{-\beta}$, with $\beta \simeq 0.6$ and 
temperature, $T \simeq 10^7$K \citep[e.g.,][]{math}. 

As the smaller galaxy moves toward the host, with a typical initial speed of 
$V_s \simeq 300$ km s$^{-1}$, \citep[appropriate for a group with a 1 keV X-ray 
temperature; e.g.,][]{jelt}, 
its gas will be subject to ram 
pressure stripping by  the CGM.  However, only the outer portions of this 
gas are likely to be directly stripped at large distances from the host where
$n_{CGM} < 10^{-3}$cm$^{-3}$, in that $M_s = 1.0 \times 10^{11} M_{\odot} \gg 
M_{IS}$, where the mass below which the simple ram pressure stripping is
efficient is given by \citep{mori}, 
\begin{eqnarray}
M_{IS} = 1.27\times 10^9 (f/0.1)^{-7/2} (n_{CGM}/10^{-4} \rm{cm}^{-3})^{7/2} 
\nonumber\\\times (V_s/10^3 \rm{km ~s}^{-1})^7 M_{\odot}~~~~~~~~~~~~~
\end{eqnarray}
A second phase of slower stripping, basically due to the Kelvin-Helmholtz 
instability, still will occur, but simulations show that its timescale should exceed $2 \times 
10^9$ yr for these parameters 
\citep{mori,roed}. 
By that time the merger should be complete if the smaller galaxy is 
fairly compact and its impact parameter, $b \lesssim  R_c$, where $R_c \simeq
10$ kpc is the galactic half-light radius \citep[e.g.,][]{bert,gonz}.

Therefore, at the point of 
the approaching galaxy's entry into the inner part of the CGM a 
significant fraction 
$\eta$,  of the infalling galaxy's gas is taken to be retained. The initial angular momentum of 
the gas associated with the galaxy to be captured is 
$\ell_i = \eta M_g b V_s$. The gravitational interaction between the stars of
the two galaxies leads to the merger through dynamical friction and tidal 
stripping; then the  separation between the remainder of the infalling galaxy and the distorted host \citep[e.g.,][]{gonz}
will be $r_f < R_c/2$. 
Any remaining orbital velocity must be less than the velocity dispersion of 
the host, so $V_f \lesssim 200$ km s$^{-1}$.  Therefore the ISM core of the 
infalling galaxy will also have shared the bulk of its angular momentum with 
the larger galaxy's gas. The decrease in the angular momentum of the 
infalling galaxy's gas will be 
\begin{equation}
\Delta \ell = \ell_i [1 - (r_f/b) (V_f/V_s)].  
\end{equation} 

 Given the impact parameter required for the merger to occur, 
the bulk of this angular momentum must be taken up by the gas of the 
  inner portion of the CGM, which will 
acquire a global orbital speed, 
\begin{equation}
\langle V_{CGM}\rangle = \Delta \ell / M_{CGM} \langle R \rangle ,
\end{equation}  
where the angle brackets denote mass weighted averages. 
For example, with $R_{max,CGM} = 20$ kpc, $M_{CGM} \simeq 2.3\times 10^9 M_{\odot}$ and 
$\langle R \rangle \simeq 12$ kpc.
Then, for $\eta = 0.8$ and $f = 0.1$ we find from Eq.\ (3) that
$\langle V_{CGM} \rangle \simeq 580$ km s$^{-1}$.  Such a very large orbital velocity 
for the gas (and not for the stars) means that enough angular mometum has
been injected into
the gas to transform the
initially  quasi-spherical CGM halo, with a thermal (sound)
speed, $V_t \simeq 470$ km s$^{-1}$, into a fat pancake of radius approaching 
40 kpc.  This expansion would take a time 
$\tau \gtrsim 40 \rm{kpc} / V_t \simeq 8 \times 10^7$ yr.  As this is much 
less than the time for the merger to be completed, we see that this merger 
mechanism does provide a viable route to the formation of a wide, flattened
pancake of gas. 
Recently, \citet{dima} 
have performed an extensive suite of simulations of mergers and fly-bys of elliptical-elliptical,
elliptical-spiral and spiral-spiral pairs, including both co- and counter-rotating
interactions at various peri-centers.  The focus of these simulations was on the amount 
and location of the enhancement of star formation, and they assume that the elliptical galaxies are devoid of the relatively cool ($\sim 10^4$K) gas that can plausibly produce stars, while they ignore the hot gas ($\sim 10^7$K) known to be present around ellipticals
and which we utilize in the above calculations.  \citet{dima} 
show that the results depend sensitively on the orientations of disks and on the maximum strength of the tidal interactions; while their simulations are not directly relevant to the situation we are treating, they do clearly show that in some cases, large fractions of the gas can be ejected far from the center of the merged galaxies, in essential agreement with our argument earlier in this paragraph.
In our rough estimate we have not explicitly considered the 
ram pressure stripping that removes more of the gas from the infalling galaxy
as it traverses the denser innermost portions of the CGM.  However, the angular 
momentum of this gas will also be absorbed by the CGM, so this should not 
drastically affect our argument.  Our calculation is merely a demonstration that for 
reasonable parameter values, not quite major mergers of elliptical galaxies can yield very expanded
hot gaseous SDs and we expect (and require) that they will actually form in only a minority of all elliptical mergers.

Observational evidence for rotation in the gas surrounding RGs comes from 
optical (H$\alpha$) spectroscopy \citep{noel}
although the
galaxies in this sample are weaker, nearby ($z < 0.023$), FR I RGs.  Unfortunately, 
the number of FR II RGs at redshifts 
low enough to allow measurements 
that could clearly detect the rotation of the hot
associated gas is small, so direct evidence for rotation in the
type of RG we are considering is not yet available.  Additional 
indirect evidence for large-scale rotation in the CGM comes from the 
comparison of observed and simulated radio source morphologies 
in nearby clusters of galaxies 
\citep{hein}. 
Detailed simulations of merging galaxies in massive, high-$z$
halos suggest that large quantities of molecular gas form
and gradually evolve into a rotational configuration and
eventually into a disky structure, in agreement with sensitive
CO line observations of quasars 
\citep{nara,cari}.

One caveat   to our merger scenario is the possibility that the  gas stripped
from the smaller galaxy will, to 
some degree, drive turbulence which might dissipate some of the angular 
momentum before it can be widely shared by the CGM.  Simulations
indicate that this effect is probably minor under most circumstances 
\citep[e.g.,][]{roed},
but no extant simulations of which we
are aware are directly applicable 
to the situation envisioned here.  It is worth noting that the presence of 
even weak magnetic fields in the CGM will substantially help stabilize the 
relatively flat boundary between the SD and the radio lobes against surface 
instabilities \citep[e.g., simulations by][]{jone}.  
The stabilizing 
role of this magnetic tension is may well be indicated by the 
commonly observed good alignment of the magnetic field within the radio 
lobes along their quasi-linear inner edges (Fig.\ 1; Table 1).

The final stage of the galactic merger, particularly once the two SMBHs 
merge, is very likely to trigger activity in the nucleus.
We earlier considered (GKBW03) how a rotating stream of the CGM gas could 
explain the  {\bf Z}-symmetry detected in a few well mapped {\bf X}-shaped 
RGs (XRGs). 
The currently favored mechanism for XRGs involves the reorientation 
of the spin axis of the SMBH in the dominant galaxy following merger with 
another substantial BH   at the core of the captured galaxy \citep[e.g.,][]{rott,zierb,merr}. 
Before the merger, a pair of 
jets is feeding one pair of radio lobes, but the merger temporarily disrupts 
the accretion disk, and thus the jets. 
The post-merger SMBH has its spin axis reoriented, largely at the expense of 
the orbital angular momentum of the captured BH \citep[e.g.,][]{merr,zierb}.
It then can reestablish new jet flows along this 
flipped axis. This type of rapid reorientation of the RGs jets
 can nicely explain the morphological and spectral 
properties of the XRGs.
Some of these new jet pairs should naturally develop a {\bf Z}-symmetry while traversing 
the CGM already set in rotation if, as is expected, the CGM and reoriented 
SMBH do not have a common rotation axis \citep{gkbw,zier}. 

In a small fraction of SMBH mergers, however, no spin-flip will occur because 
the orbital axis of the infalling BH happens to be close to the original 
spin axis of the central engine.  
For such cases the axis of the flattened CGM disk will also be close to both 
the old and new jet axes; therefore, the jets will continue to propagate 
undeflected.  Such undeflected ``naked jets'' are visible, for example, in 
the radio sources 3C 28 and 3C 153 (DRAGN atlas: www.jb.man.ac.uk/atlas). 
The blocking of the backflow by the thick and now flattened pancake of CGM 
would give rise to the strip-like emission gap between the radio lobes, i.e.,
the SD.  While the rotation of the CGM disk would have no major influence on 
the propagation of the stiff jets, the lobe plasma within the region of the
rotating CGM is likely to be peeled off of the jets and mixed into the CGM,
thereby causing a drastic fading of the lobe radio emission within the disk.  
Such residual radio emission within the SD could sometimes still be
detected; some examples are the RGs J1137$+$613 \citep{lara01} and 3C 192,
where a deeper map at lower frequency 
shows faint radio emission coinciding with the SD \citep{baum}. 

Note that in our scenario, the number of SD sources should be 
significantly smaller than that of the XRGs, as they both 
arise from SMBH mergers, but the former requires a preferred orientation.  
Until recently, however,  RGs showing SDs and {\bf X}-shaped lobes were known 
in comparable numbers, 
thereby posing an apparent inconsistency. However, a recent analysis of the 
FIRST \citep{beck} 
survey has led to the discovery of 
over 100 new XRG candidates \citep{cheu}, 
so the relative numbers of XRGs 
and SDs no longer seems to be inconsistent with our proposal.

To understand the asymmetrical placement of the host galaxy with respect 
to the center of the emission gap in the sources we have focused on in this
paper, we require the presence of a modest component of the motion
of the host galaxy along the direction of the jets (GKW00).  
This would be $\sim 100$ km s$^{-1}$ to cover the typical offset distance of
$\sim 10$ kpc in a RG lifetime of $\sim 10^8$yr \citep[e.g.,][]{bara}.
Some additional
movement of the host, roughly along the
radio axis, arises from the linear momentum imparted
by the incoming galaxy at the time of merging.  However,
for the mass ratio (1:5) 
we have taken as fiducial, 
and for a component of the galaxy's velocity along the axial direction
of $V_k = 3^{-1/2}~V_s \simeq 170$ km s$^{-1}$, this kick amounts to 
only $\sim 30$ km s$^{-1}$.  The merger is therefore unlikely to cause the
bulk of the asymmetry, which must be provided by random motion of the
host galaxy; however, it certainly can contribute to it. 

In any event, the present finding that the lobe pair is more symmetric 
relative to the SD than to the host galaxy distinctly favors the scenario
in which motion of the galaxy relative to the SD material, during the lifespan 
of radio activity, makes a substantial contribution to the observed 
lobe-length asymmetry measured relative to the radio core (GKW00). 
Here it is worth reiterating that although orientation effects can 
significantly distort the observed properties of most samples of RGs
\citep[e.g.,][]{hump,gkw05},
they are expected to be minimal for RGs 
exhibiting signatures of superdisks (Sect.\ 1). 
An examination of asymmetry in the kinematics 
of quiescent gaseous halos associated with powerful RGs and 
their radio parameters, such as the brightness of jets, hotspots and the
spectral index, led \citet{hump}
to conclude that while the 
statistically significant correlations found among these parameters 
can be reasonably explained in terms of orientation effects, the predicted 
pattern of radio lobe-length asymmetry is, however, not observed. They 
attribute this apparent discrepancy to the primary influence of environmental 
asymmetry on the lobe length attained. In our picture, this {\it decoupling} 
of the lobe-length asymmetry from the various orientation induced correlations
can be understood in terms of a large contribution to the lobe-length
asymmetry coming from the galaxy motion (which is orientation independent),
as mentioned above.


\section{Conclusions}

We have highlighted a particularly interesting and intriguing class of RGs 
which not only shows an SD type morphology but also the host 
galaxy is seen close to the edge of the SD and 
is thus grossly offset from the midplane of the SD. This morphological 
asymmetry poses a serious difficulty for existing SD models, particularly 
when the superdisk is found in RGs at $z \lesssim 1$, where a significant 
amount of hot gas surrounding the host is usually present.
We have shown that the two radio lobes extend significantly 
more symmetrically about the SD midplane than they extend 
relative to  the host galaxy.

We have proposed a scenario whereby gaseous SDs acquire their form through 
the injection of angular momentum from the gas belonging to a galaxy whose eventual
merger with the massive elliptical host triggers the jets
responsible for the RG. The gradual transfer of the angular momentum into 
the hot CGM of the host elliptical flattens its CGM into a fat-pancake 
shaped superdisk, which can account for the strip-like emission gaps 
observed between the radio lobes of such RGs. In this picture, the grossly 
off-centered location of the host (relative to the central plane of the SD) 
is relatively straightforward to understand in terms of the motion of 
the host galaxy during the period of jet activity, partly assisted by a linear momentum kick imparted by the merged 
galaxy. A corollary to this picture is that the SD material (which appears
to be ``docking the tails'' at least in these RGs 
\citep[cf.,][]{jenk} 
is not gravitationally bound to the radio-loud elliptical. 
Interestingly, this scenario also provides a possible physical link between 
the SDs  and the XRGs that are widely believed to manifest 
mergers of two galaxies containing SMBHs.

The merger is likely to launch a pair of jets and if the host had already
been recently active, a double-double radio structure can arise, as
seen in three RGs in the present sample (0035$+$130, 0114$-$476, and
1155$+$256). It is interesting that  for between 10 and 13 of these 16 RGs, 
the brighter of the two outer hotspots is associated with the radio lobe 
adjoining the host galaxy.  This indicates a diminishing of a jet's
radio emitting potential following a long passage through the denser 
thermal material associated with the SD, as compared to a jet that mainly 
traverses the (light) relativistic plasma within a radio lobe.
Perhaps a striking manifestation of this is seen in the RGs 0106$+$729 and 
1939$+$605  where the radio jet appears to turn dissipative after 
propagating through the SD.  No such dissipation is observed,  at least
in the present
sample, for the jets that propagate mainly through the radio lobe. This
calls for a follow-up study once a significantly larger sample of highly
asymmetric superdisks becomes available.

We intend to investigate the realm of applicability of this model 
through simulations of the merger of two  massive ellipticals with 
substantial circumgalactic gas.

\section*{Acknowledgments}

We thank Santosh Joshi for assistance in compiling the present sample,
Scott Tremaine for correspondence and Drs.\ Baum, Leahy and Owen for
permission to republish maps.  We appreciate the careful comments of the
anonymous referee that have led to a significant improvement in the presentation
of these results.  This research has made use of the NASA/IPAC 
Extragalactic Database (NED) which is operated by the Jet Propulsion Laboratory, 
under contract with NASA. PJW is grateful for  hospitality at NCRA, 
Princeton University and the Institute for Advanced Study. 

\end{document}